\documentclass[final,5p,times]{elsarticle}

\usepackage{amsmath}
\usepackage{amssymb} 
\usepackage{fancyhdr}
\usepackage{multicol}
\usepackage[table]{xcolor}
\usepackage{caption}
\usepackage{subcaption}
\usepackage{comment}
\usepackage{multirow}
\usepackage[T1]{fontenc}
\usepackage{mathptmx}
\usepackage{xfrac}
\usepackage{wrapfig}
\usepackage[squaren, Gray, cdot]{SIunits}
\usepackage{array}
\RequirePackage[width=.90\linewidth]{caption}
\usepackage{pdfpages}
\usepackage{appendix}
\usepackage{booktabs}
\usepackage[nobottomtitles*]{titlesec}
\usepackage{lineno}
\usepackage{hhline}
\usepackage{tabularx}
\usepackage{booktabs}
\usepackage{xparse}
\usepackage{ulem}
\usepackage{color,soul}
\NewExpandableDocumentCommand\mcx{O{1}m}
    {\multicolumn{#1}{>{\Centering\small\bfseries\hsize=#1\hsize}X}{#2}}
\usepackage{ragged2e}

\DeclareMathAlphabet{\mathcal}{OMS}{cmsy}{m}{n} 
\usepackage[pdfencoding=auto, psdextra, hidelinks]{hyperref}

\newcommand*{\fullref}[1]{\hyperref[{#1}]{\textit{\ref*{#1}-\nameref*{#1}}}}
\newcommand{\cevns}{CE$\nu$NS}

\newcommand{\nucleus}{\textsc{Nucleus} }
\newcommand{\ricochet}{\textsc{Ricochet} }
\newcommand{\Geant}{\textsc{Geant 4} }

\journal{Elsevier}

\begin{document}
\begin{frontmatter}



\title{Prototyping a High Purity Germanium cryogenic veto system for a bolometric detection experiment}


\author[1]{C.~Goupy\corref{cor1}}
\author[2]{S.~Marnieros}
\author[1]{B.~Mauri\fnref{now}}
\author[1]{C.~Nones}
\author[1]{M.~Vivier}
\cortext[cor1]{chloe.goupy@cea.fr}
\fntext[now]{Now at: Max-Planck-Institut für Physik, München}
\affiliation[1]{organization={IRFU, CEA, Université Paris-Saclay},
            postcode={91191}, 
            city={Gif-sur-Yvette},
            country={France}}
\affiliation[2]{organization={CNRS/IN2P3, IJCLab,  Université Paris-Saclay},
            postcode={91405}, 
            city={Orsay},
            country={France}}
\begin{abstract}
The use of High Purity Germanium detectors operated in ionization mode at cryogenic temperatures is investigated as an external background mitigation solution for bolometers used in rare-event search experiments. A simple experimental setup, running a 52-g $\mathrm{Li_2WO_4}$ bolometer sandwiched in-between two 2-cm thick High Purity Germanium cylindrical detectors in a dry cryostat, shows promising rejection to environmental gammas and atmospheric muons backgrounds. The acquired data are used together with a Monte Carlo simulation of the setup to extract the main contributions to the external backgrounds expected in an above ground experiment, such as e.g.~current and future experimental efforts targeting the detection of coherent elastic neutrino-nucleus scattering at reactor facilities. Based on all these results, a $\mathrm{4\pi}$ coverage similar veto system achieving a $\mathcal{O}$(10 keV) energy threshold is expected to achieve a $\mathrm{\gtrsim}$\,73\,\% and a $\mathrm{\gtrsim}$\,92\,\% rejection power for gamma-like and muon-like events, respectively.
\end{abstract}
%


\end{frontmatter}



\section{Introduction}
Over the past decades, the steady progresses achieved in the development of cryogenic detectors for particle and nuclear physics widespread their use in rare-event search experiments. The high energy resolution, the large sensitivity range and the possibility to implement multiple read-out strategies using heat, scintillation, and ionization signals make them extremely attractive and suitable devices for e.g.~dark matter searches \cite{Cresst, EdelweissTES, SuperCDMS} or searches for neutrino-less double beta decay \cite{DoubleBeta}.~Although these detectors can already be very effective in identifying and discriminating unwanted sources of background, the quest for ever-higher sensitivities in the future generation of bolometric detection experiments call for innovative background rejection strategies pursuing for the highest efficiencies. This feature is especially relevant for the shortcoming generation of experiments aiming at studying coherent elastic neutrino-nucleus scattering (\cevns) at reactors using cryogenic detectors, such as e.g. \nucleus \cite{NUCLEUS2019} or \ricochet \cite{Ricochet}. The detection of \cevns{}-induced nuclear recoils down to sub-keV energies still requires effective background rejection techniques in these uncharted very low energy regimes, and in which the ionization or scintillation yield degrades, making an effective background mitigation strategy solely based on a dual-readout scheme very challenging to achieve. To this end, this article investigates for the first time the feasibility of using a compact cryogenic veto system in close vicinity to a target cryogenic detector for mitigating external particle backgrounds.~The use of High Purity Germanium (HPGe) ionization detectors, which are known to have a rather low $\mathcal{O}$(keV) energy threshold and fast timing response, is here considered. 

The article is structured as followed: section \ref{sec:Setup} details the experimental setup while section \ref{sec:Exp_results} reports on the characterization of each detector response and the ability of the HPGe detector to reject external events in the target cryogenic detector through an anti-coincidence analysis. This experimental work is completed by a \Geant simulation study described in section \ref{sec:Simulation} to separate the various particle sources contributing to each detector single event spectrum and to confirm the measured background rejection power of the setup. Finally, section \ref{sec:conclusions} summarizes the main results of this work and opens up possible improvements of this setup in view of reaching the highest external background rejection efficiencies.
\section{Experimental setup}\label{sec:Setup}
\begin{figure*}[ht!]
    \centering
    \includegraphics[width=0.8\textwidth]{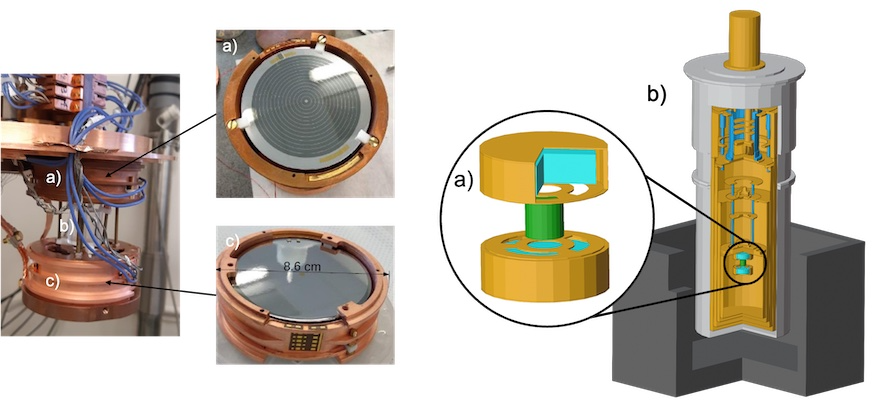}
    \captionsetup[figure]{width=\linewidth}
    \captionof{figure}{\textit{(Left)} Photographs of the experimental setup. \textbf{a)} TOP Ge crystal, equipped with concentric ring ID electrodes. \textbf{b)} LWO crystal, at the centre. \textbf{c)} BOT Ge crystal equipped with planar electrodes. The mechanical structure holding the detectors and the blue low noise cables connected to the HPGe electrodes are also visible.\newline
    \textit{(Right)} Cut view of the simulated \Geant geometry (see section \ref{sec:Simulation} for further details). \textbf{a)} Focus on the detection setup. The HPGe crystals are colored in light blue, whereas the central LWO crystal is green-colored. The copper housings are represented in gold. \textbf{b)} The simulated geometry of the cryostat with the surrounding Pb shielding (dark gray). Main materials are copper (gold), aluminum (gray) and stainless steel (blue).}
    \label{fig:COVProto}
\end{figure*}

The experimental setup is pictured and sketched on figure \ref{fig:COVProto}.~It consists of two cylindrical $\mathrm{\sim}$\,400\,g HPGe crystals of 70\,mm diameter and 20\,mm height, each placed $\simeq$\,40\,mm above and below a 25\,mm diameter and 25\,mm height 51.7\,g cylindrical $\mathrm{Li_2WO_4}$ crystal \cite{BASKET,LWO_Growth}. The solid angle coverage provided by this HPGe veto arrangement is $\sim 2\,\pi$\,sr.

The two HPGe crystals were operated in ionization mode using Aluminum (Al) electrodes evaporated on their respective top and bottom surfaces to collect the electron/hole charge carriers.~These 200-nm thick electrodes were deposited by electron beam evaporation after an Argon (Ar) cleaning and passivation step using a 30-nm thick amorphous Ge:H layer \cite{Evap}. The bottom crystal (BOT Ge) is equipped with planar electrodes. The top crystal (TOP Ge), which was previously used in the \textsc{Edelweiss} experiment \cite{Edelweiss2010}, is equipped with ring-shaped electrodes (ID electrodes). These electrodes were specially designed to discriminate between surface and bulk events.~Because this feature was unnecessary in the present work, the ring electrodes were serially connected together to mimic as closely as possible the electric field resulting from polarized planar electrodes such as those equipping the BOT Ge.~The two HPGe crystals were housed within 3-mm thick copper boxes and secured using PTFE (Teflon) holders. The thermal link and the electrode signal transmission were ensured by bonding the crystals using a Kapton printed circuit board glued onto the copper box. The copper housing lids facing the LWO detector were mechanically designed to hold the LWO crystal using PTFE pieces. The HPGe detectors were polarized with a 5~V/cm electric field by biasing one electrode at +10~V and grounding the opposite one. An infra-red LED (EOLS-1650-199 from EPIGAP) was glued to the BOT Ge housing in order to regenerate the HPGe crystals \cite{regen_Ge}. The two crystals were daily illuminated for ten minutes, after which one hour was necessary to stabilize the cryostat back to base temperature. An opened and collimated $\mathrm{^{241}}$Am source was mounted underneath the BOT Ge detector. The source was covered by a thing copper tape to stop $\mathrm{alpha}$ particles.

The LWO crystal was operated as a bolometer and read out with a Germanium Neutron-Transmutation-Doped (Germanium NTD) thermal sensor \cite{NTD}. It was used as a target detector to test the identification and the rejection power of external particles using the HPGe ionization detectors operated in anti-coincidence mode. The NTD sensor was biased with a 1-kHz alternating constant current ($\pm$ I$_\text{NTD}$) in order to improve the signal-to-noise ratio \cite{NTD-Modulation}. 
The full mechanical assembly was wrapped in a copper foil to shield against possible infra-red radiations.~The experimental setup was operated at a temperature of $\mathrm{\sim}$18 mK in the so-called "Actuator" cryostat at Laboratoire de Physique des Deux Infinis Irène Joliot-Curie (IJCLab, Orsay), which is a custom-made pulse-tube dilution refrigerator embedded in a single vacuum chamber (see e.g. \cite{thesis_beatrice} for further details). No vibration decoupling systems were used.

\begin{figure}[ht!]
    \centering
    \includegraphics[width=0.8\linewidth]{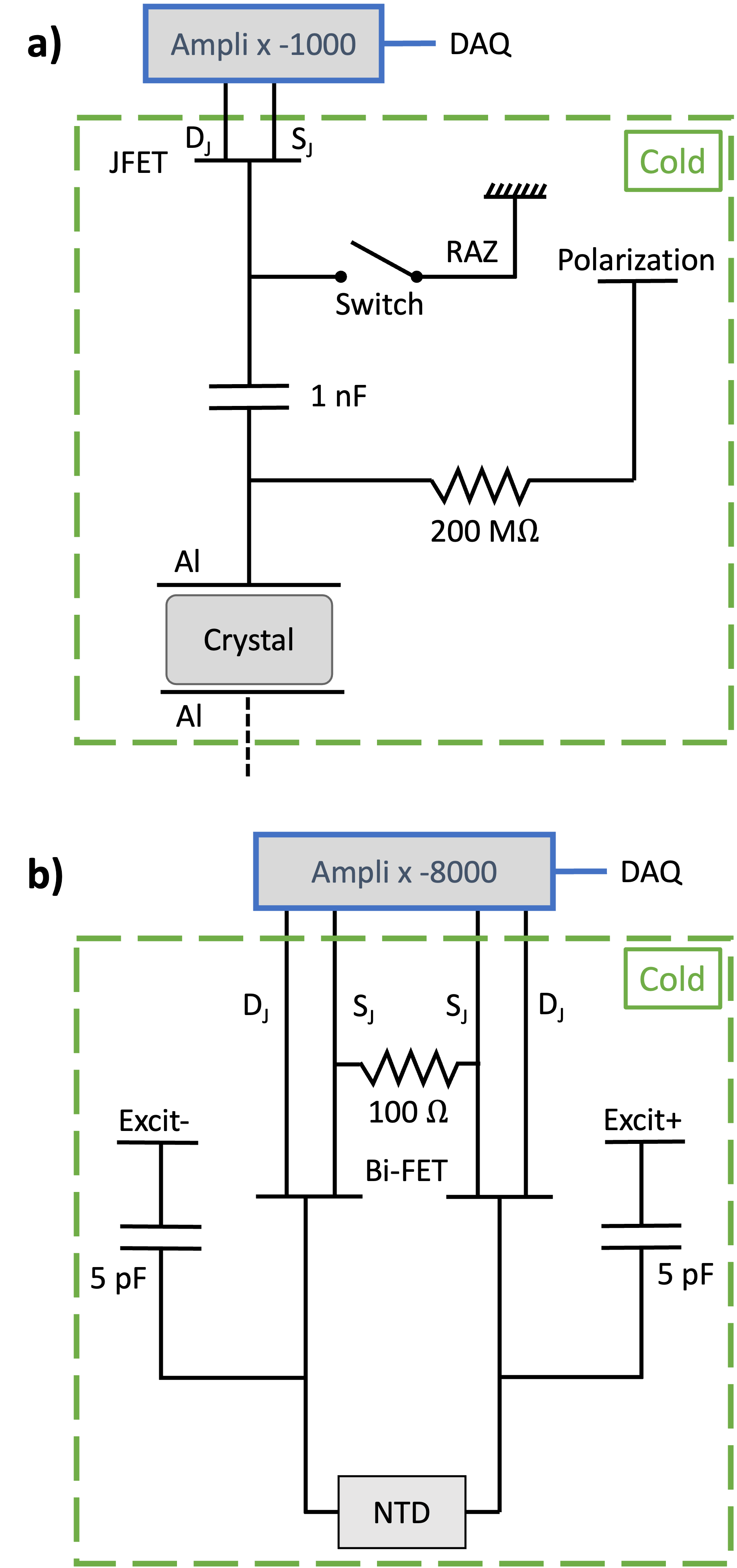}
    \captionsetup[figure]{width=\linewidth}
    \captionof{figure}{Cold electronic schemes for \textbf{a)} the HPGe detectors and \textbf{b)} the LWO bolometer. Not represented in the figure, the NTD read-out features as the HPGe read-out, a switch to discharge the bi-FETs when necessary. The blue boxes contain the warm post-amplification stage (also featuring anti-aliasing low-pass filters). 
    }
    \label{fig:elec_scheme}
\end{figure}

The detectors were connected with low noise stainless steel cables. These coaxial-cables especially feature a carbon coating between the core and the shielding to dissipate the accumulation of electrostatic charges induced by mechanical vibrations, hence preserving the signal integrity and reducing microphonic noise.~The electronic operation and acquisition scheme is detailed on figure \ref{fig:elec_scheme} for both the HPGe ionization detectors and for the LWO bolometer. A first signal amplification step is performed in a cold stage using Junction Field Effect Transistors (IFN860 bi-JFETs). The output voltage signals are then filtered and further amplified through voltage amplifiers in a second electronic stage operated at room temperature.~Signal acquisition was done with a commercial NI6366 X Series Data Acquisition system from National Instruments.~Output signals were continuously sampled and read out at a 100~kHz frequency.
Data were collected in two experimental configurations: (i) without the use (10 hours of acquisition time) and (ii) with the use (40 hours of acquisition time) of an external lead (Pb) shielding (see right panel of figure \ref{fig:COVProto}) placed around the cryostat. The purpose of this Pb shielding was to reduce the pile-up of events in the slow LWO detector, which features pulses of 0.5-1\,s decay times and a rate of $\sim$ 5\,Hz. Adding the Pb shielding reduced this rate by a factor 2.5. The lead shielding was made of three 5-cm thick brick walls and one partial wall of 10-cm thickness, hence partially covering the detector setup. During those two data acquisition runs, the detector setup was exposed \sout{punctually} for a few hours, to a $\mathrm{^{232}Th}$ source for energy calibration and a $\mathrm{^{252}Cf}$ neutron source for quenching measurements. 

\section{Experimental results}\label{sec:Exp_results}
The operation of HPGe ionization detectors as an efficient veto system for a bolometric target detector imposes good energy and time reconstruction together with a high pulse selection efficiency down to the lowest possible threshold. This latter requirement is particularly relevant for veto detectors operated at surface, for which event 
rates can lead to significant pile-up probability. While a good timing resolution allows to keep accidental coincidences at acceptable rates, precise energy determination may prove to be important to better identify potential background sources present in the experimental setup environment.~To meet these requirements, a fast and efficient pulse identification algorithm was developed. It takes advantage of the large difference between the detector pulse rise and decay time components, then making it almost insensitive to the presence of pile-up events. In this situation, the information about the deposited energy is mostly contained in the rise time region of the pulse. Event triggering and energy determination can then be performed in a rather small time window, hence keeping under control possible pulse distortions from pile-up effects. The present event selection and reconstruction algorithms have been applied to both the HPGe veto detectors and the LWO bolometer data streams. They are described in the following subsections.

\subsection{Event selection and reconstruction} \label{sec:Analysis_method}
\begin{figure}[ht!]
    \centering  
    \centering
    \includegraphics[width=0.95\linewidth]{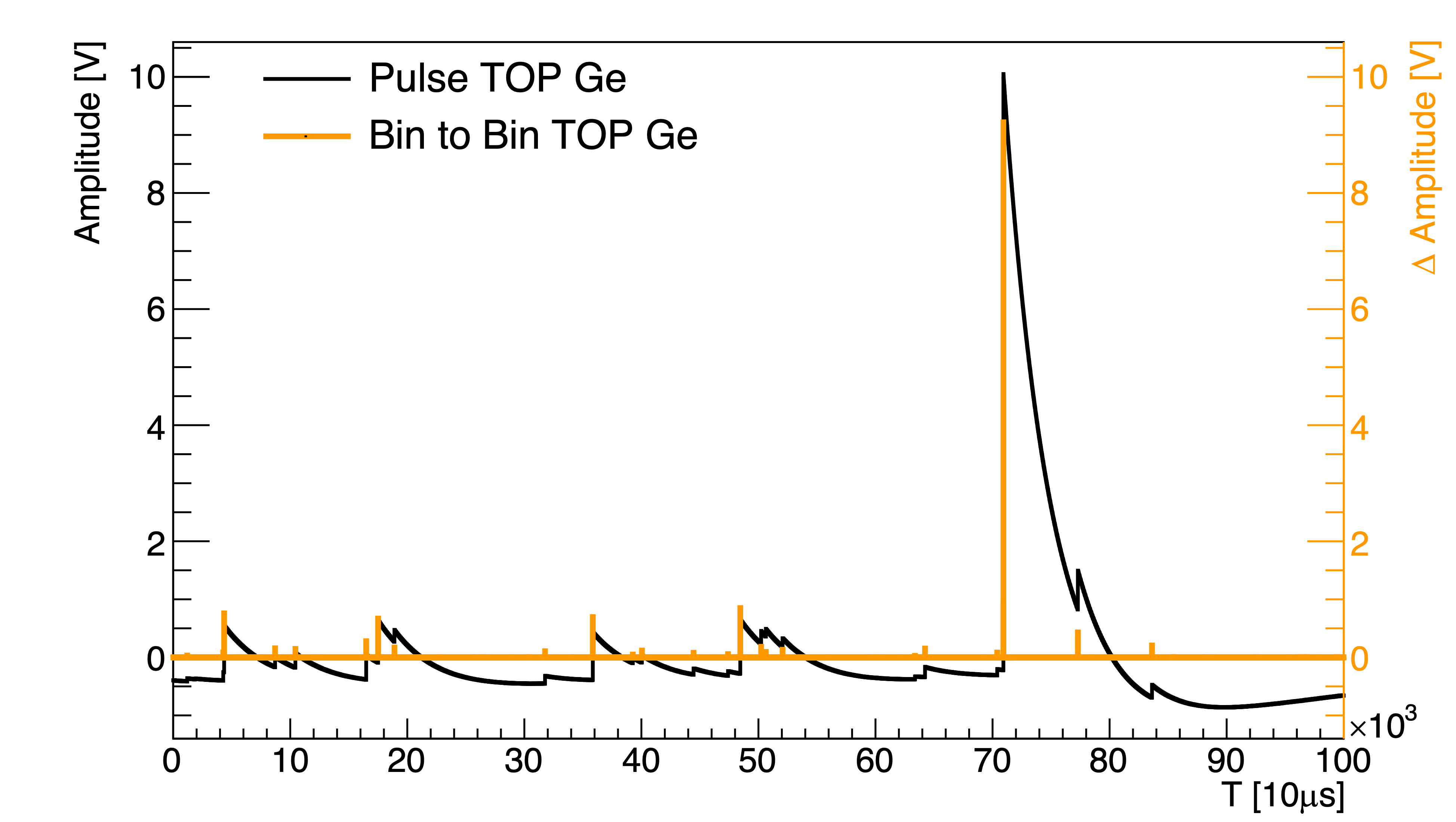}
    \captionsetup[figure]{width=\linewidth}
    \captionof{figure}{Example of recorded ionization pulses over a 1-s time period for the TOP Ge detector (black line). The orange line depicts the bin-to-bin differentiation of this data stream. This differentiation method both allows to recover a flat baseline and to clearly identify events, owing to the large difference between the rise time and the decay time of the original pulses.}
    \label{fig:Ge_Pulses}
\end{figure}

\begin{figure}[ht!]
    \centering
    \includegraphics[width=0.95\linewidth]{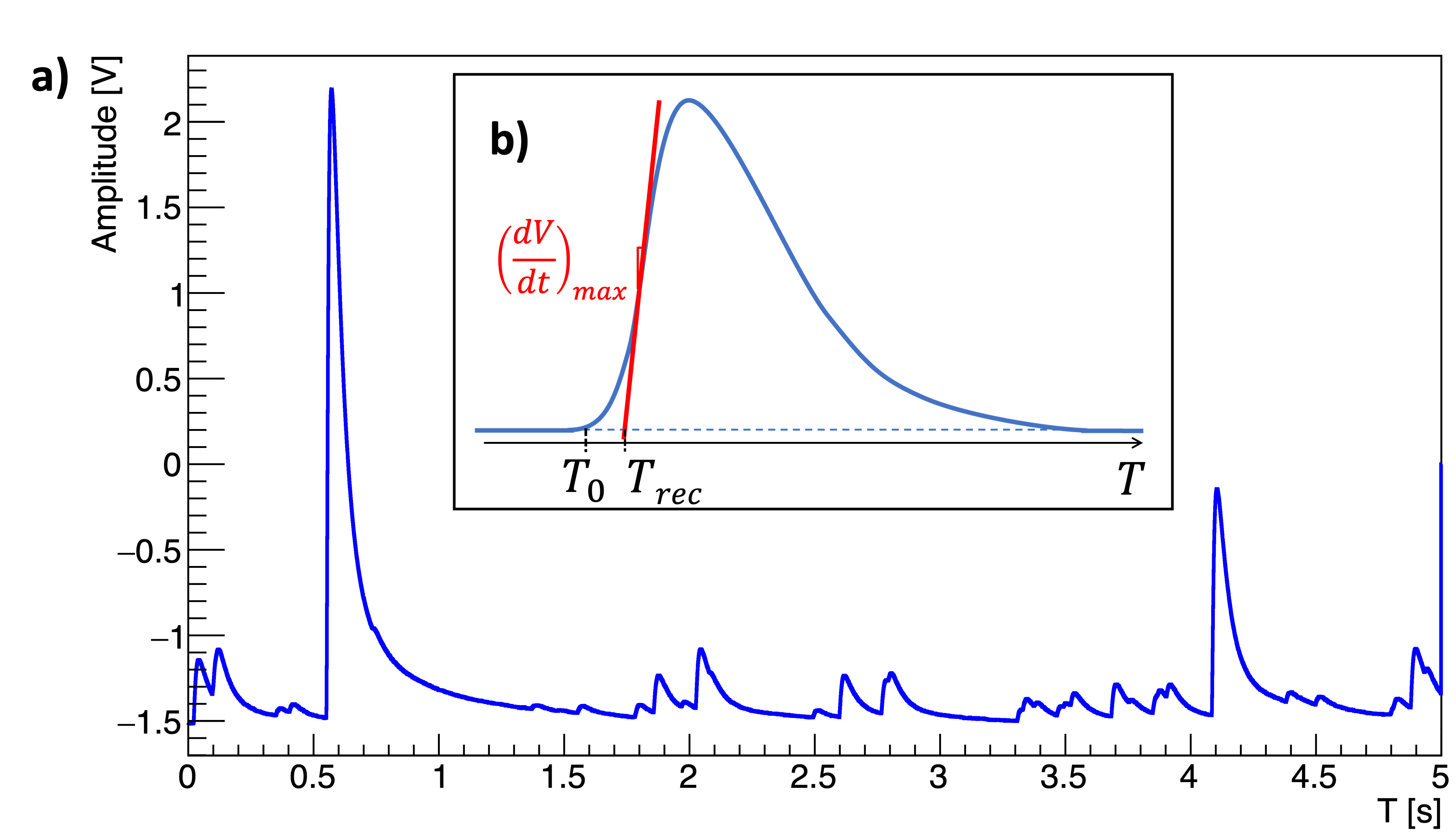}
    \captionsetup[figure]{width=\linewidth}
    \captionof{figure}{Typical pulses from the LWO NTD detector \textbf{(a)}. The method used to estimate the time T$_{rec}$ of a selected pulse is pictured in the inset (\textbf{b}). See text for further details.}
    \label{fig:NTD_Pulses} 
\end{figure}

The event selection and reconstruction take advantage of the HPGe and LWO detector pulse features, which exhibit a fast rise time with respect to their decay time. The amplified signals from the Ge detectors were measured to typically have an average rise time less than 30 $\mu$s and a decay time typically in the 30 to 50 ms range (see figure \ref{fig:Ge_Pulses}). This feature is also present in the NTD sensor, which delivers much slower signals with rise times of $\simeq$ 25 ms and decay times up to 1 s (see figure \ref{fig:NTD_Pulses}). As illustrated by figure \ref{fig:Ge_Pulses}, the bin-to-bin differentiation of such signals allows to clearly identify physical pulses above a flat baseline centered around zero. Events are thus triggered by setting an appropriate threshold above the measured baseline fluctuations. The energy associated to each detected pulse is then reconstructed by summing the amplitudes of the discretely differentiated signals over a time window covering the pulse leading edge region (i.e. non-zero derivative), and after applying an energy calibration factor. Such a procedure is equivalent to measuring the full amplitude rise of the detector signal.

For the three detectors, data were originally acquired with a 100 kHz sampling frequency. For event reconstruction, a downsampling procedure was applied to the data streams at the analysis level to reduce baseline fluctuations and to improve energy resolution. It was performed by merging and averaging the content of $n$ consecutive samples, with $n$ being the downsampling factor. The HPGe detector streams were downsampled by a factor 2, leading to a 50 kHz sampling rate. The streams of the much slower LWO detector were downsampled by a factor 250, leading to a sampling rate of 400 Hz. In both cases, this choice resulted from a trade-off between energy and time resolution. In order to reject possible spurious events in the LWO detector, pulses with less than 10-ms rise time were rejected.

The energy calibration of the three detectors was performed identifying intense gamma-ray lines originating from the decay of primordial radionuclides naturally present in the surrounding environment and from the $\mathrm{^{232}Th}$ source to which the setup has been exposed to: $\mathrm{^{208}Tl}$ ($\mathrm{^{232}Th}$ decay chain) at 510.7\,keV, 583.2\,keV and 2614.5\,keV, $\mathrm{e^{+}e^{-}}$ annihilation at 511\,keV, $\mathrm{^{214}Bi}$ ($\mathrm{^{238}U}$ decay chain) at 609.3\,keV, $\mathrm{^{228}Ac}$ ($\mathrm{^{232}Th}$ decay chain) at 911.2\,keV and 969.0\,keV, and $\mathrm{^{40}K}$ at 1460.8\,keV. An additional 59.5\,keV gamma-ray line from the $\mathrm{^{241}Am}$ source was also used for the BOT Ge detector. The next subsections report on the characterization and performances of each detector. Results are summarized in table \ref{tab:RunSummary}.
\begin{table*}[ht!]
    \centering
    \renewcommand{\arraystretch}{1.2}
    \begin{tabularx}{0.85\linewidth}{@{}@{\extracolsep{\fill}} cccccc@{}}
        \toprule
        \textbf{Detector} & \textbf{Calibration} & \textbf{Resolution} & \textbf{Mean of baseline} & \textbf{Signal} & \textbf{Rate (inside} \\
         & [$\mu$V/keV$_\text{ee}$] & \textbf{@1461 keV$_\text{ee}$} [\%] & \textbf{r.m.s.} [keV$_\text{ee}$] & \textbf{rise time} & \textbf{Pb)} [Hz] \\
        \midrule
        TOP Ge & 651.5 $\pm$ 0.1 & 0.6 &1.8  & < 10\,$\mu$s & $\simeq$\,8 \\
        BOT Ge & 949.9 $\pm$ 0.4 & 0.5 & 1.1 & < 10\,$\mu$s & $\simeq$\,15 \\
        LWO & 236.1 $\pm$ 0.2 & 0.8 & 5.7 & $\simeq$\,25\,ms & $\simeq$\,2 \\
        \bottomrule
    \end{tabularx}
    \captionsetup[table]{width=0.95\linewidth}
    \captionof{table}{Summary table of measured experimental parameters for the three detectors.}
    \label{tab:RunSummary}
\end{table*}
\begin{figure*}[ht!]
    \centering  
    \begin{minipage}{0.5\linewidth}
        \centering
        \includegraphics[width=\textwidth]{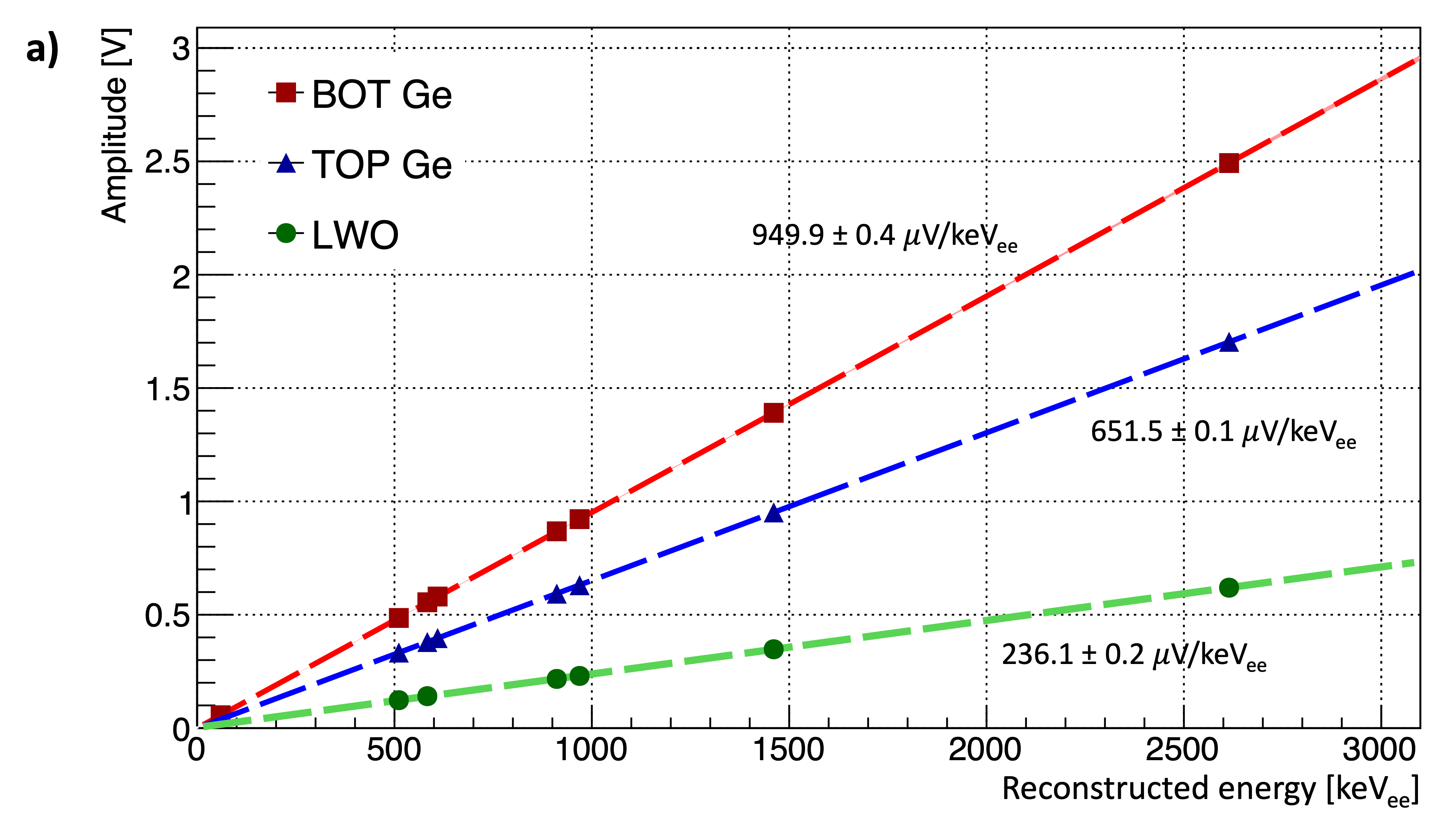}
    \end{minipage}\hfill
    \begin{minipage}{0.5\linewidth}
        \centering
        \includegraphics[width=\textwidth]{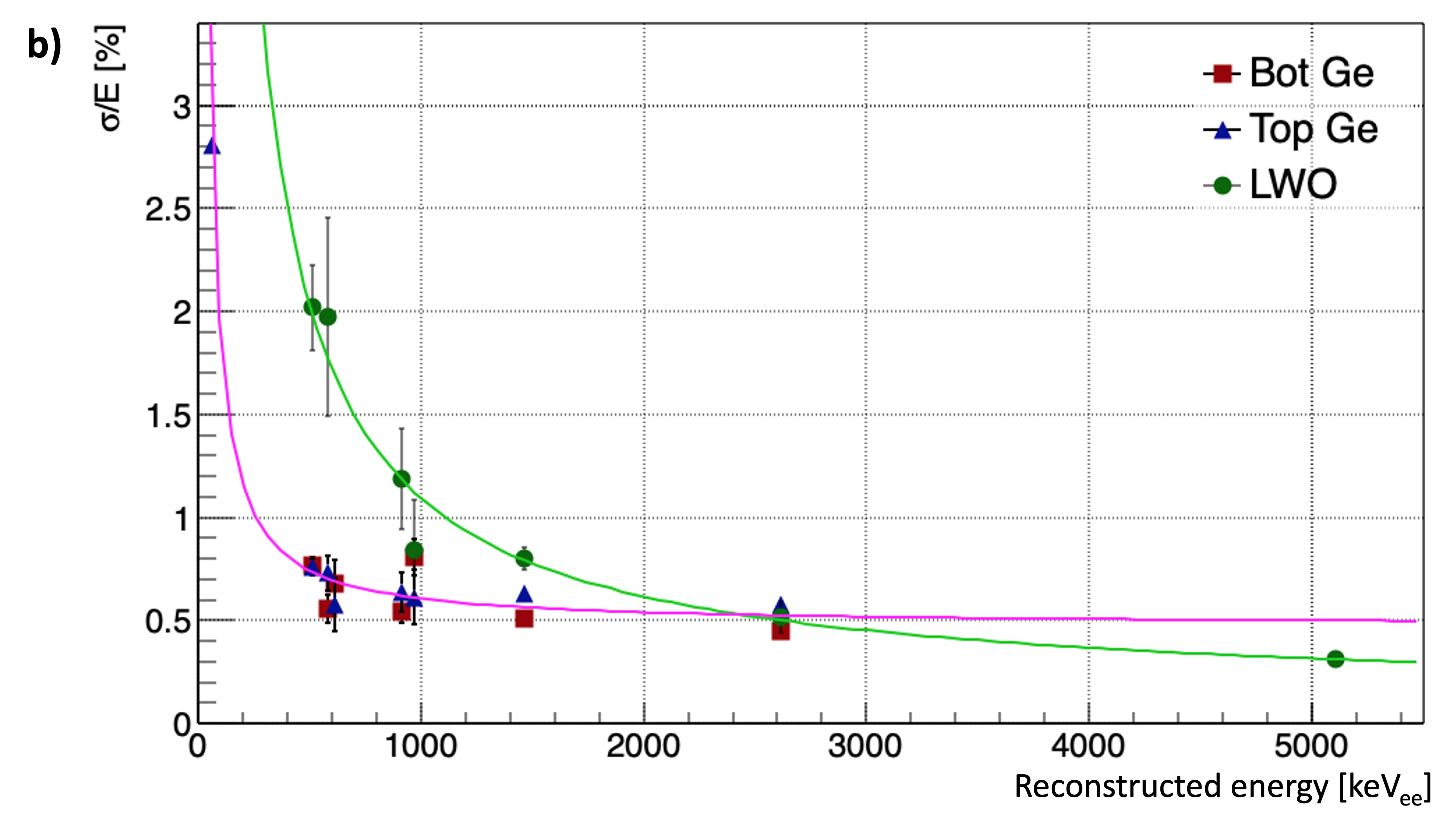}
    \end{minipage}
    \captionsetup[figure]{width=\linewidth}
    \captionof{figure}{Energy response of the HPGe and LWO detectors. \textbf{(a)} Energy calibration performed with gamma-ray lines, and fitted with a linear function. \textbf{(b)} Measured resolution as a function of energy. The magenta (resp. green) solid line shows the fit of a typical resolution function to the combined BOT and TOP Ge (resp. LWO) data. See text for further details.}
    \label{fig:Ge_Calib}
\end{figure*}

\subsection{HPGe detector results}
Figure \ref{fig:Ge_Calib}(a) shows the calibration curves obtained by fitting a linear function to the identified gamma-ray peaks. Taking into account the gain of 1000 provided by the amplification stage, the BOT Ge detector featuring planar electrodes is found to have a $\mathrm{949.9 \pm 0.4}$\,nV/keV$_{ee}$ sensitivity while a $\mathrm{651.5 \pm 0.1}$\,nV/keV$_{ee}$ sensitivity is measured for the TOP Ge detector having the ring-shaped electrodes. Non-linearities in the energy scale below 3 MeV$_{ee}$ are found to be negligible.
\begin{figure*}[ht!]
    \centering
    \includegraphics[width=0.85\textwidth]{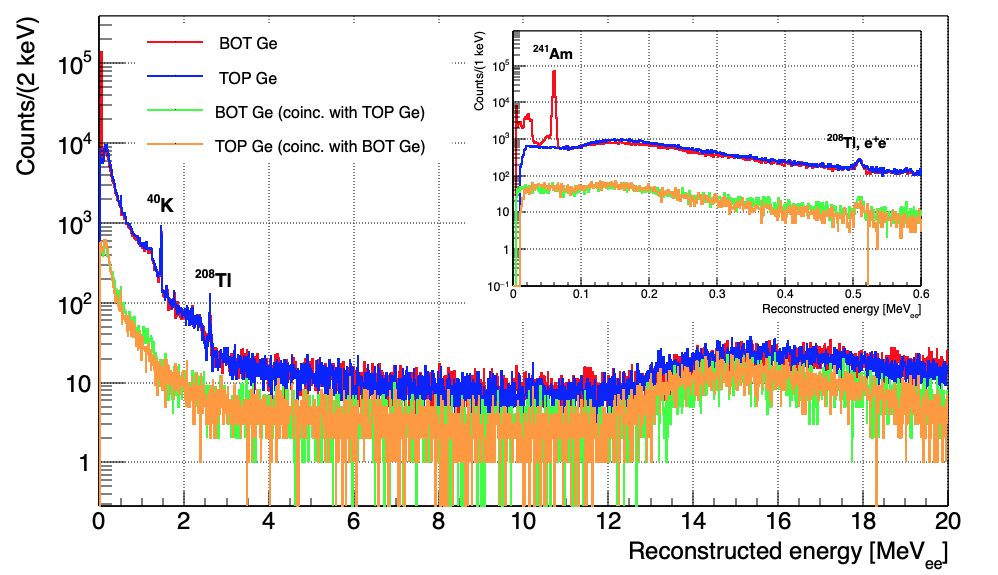}
    \caption{Energy spectra of BOT and TOP Ge detectors. The low energy range below 600\,keV$_\text{ee}$ is shown in the inset. The red and blue histograms are the triggered spectra in the BOT and TOP Ge detectors, respectively. The orange (resp. green) spectrum corresponds to the energy of the events selected in the TOP (resp. BOT) Ge detector in coincidence with the BOT (resp. TOP) Ge detector. }
    \label{fig:Ge_ESpectra_COV}
\end{figure*}

Relative energy resolutions measured at the previously mentioned gamma-ray line energies are displayed on figure \ref{fig:Ge_Calib}(b). They are found to be better than 1\,\% above 400\,keV$_\text{ee}$ and are very similar for both detectors. They are well described by the characteristic energy dependence as expected from Ge semiconductor detectors \cite{knoll}: 
\begin{equation}
    \mathrm{\frac{\sigma}{E} = \sqrt{\frac{A}{E} + \frac{B}{E^2} + C},}
    \label{eq:resolution_HPGe}
\end{equation}
where the parameters A, B and C are respectively the contributions of the charge carrier number statistical fluctuations, of the electronic noise and of the charge collection efficiency to the resolution. A fit of equation \ref{eq:resolution_HPGe} to the combined TOP and BOT Ge resolution data points gives $A =\mathrm{(1.2 \pm 0.3)\times 10^{-2}}$\,keV$_\text{ee}$, $B=\mathrm{1.9 \pm 0.2\,keV_{ee}^{2}}$ and  C = $\mathrm{(2.2 \pm 0.2)\times10^{-5}}$ (with a $\mathrm{\chi^2/\text{NDF} = 3.3}$), showing that electronic noise mostly contributes to the energy resolution at low energies ($\mathrm{\lesssim 100}$\,keV$_\text{ee}$) and that charge collection efficiency dominate at higher energies ($\mathrm{\gtrsim 1}$\,MeV$_\text{ee}$).

Electronic noise contribution for each of the HPGe detector was independently characterized by sampling the baseline fluctuations during each data run. The mean of the distribution of the baseline r.m.s. values were measured to be 1.1\,keV$_\text{ee}$ and 1.8\,keV$_\text{ee}$ for the BOT and TOP Ge detectors, respectively. These measurements come in very good agreement with the previously mentioned fit results (see figure \ref{fig:Ge_Calib} (b)). The larger value, in keV$_{ee}$ units, observed for the TOP Ge detector is mostly due to its smaller sensitivity. Some spurious periodic pulses were also identified and always reconstructed with energies below 15\,keV$_\text{ee}$. After investigation, those were attributed to cross-talk effects originating from the 1-kHz modulation current flowing through the LWO NTD sensor. Since the sampling and modulation clocks were synchronous, this cross-talk noise could be mitigated by removing very low energy pulses that were identified in-time with the leading edges of the NTD current modulation signal. This requirement introduced a small inefficiency of at most a few percents in the selection of events below 15\,keV$_\text{ee}$.

With the noise treatment described above, detection thresholds of 5\,keV$_\text{ee}$ and 10\,keV$_\text{ee}$ were applied to select and to reconstruct events in the BOT and TOP Ge detectors, respectively. The corresponding counting rates, integrated over the full energy spectra up to 20\,MeV$_\text{ee}$, were measured to be about 15 and 8 counts/s, respectively, during a typical run using the Pb shielding. The difference in rate between the two detectors can be mostly accounted for by the activity of the $\mathrm{^{241}Am}$ source. Figure \ref{fig:Ge_ESpectra_COV} further illustrates this difference, where the energy spectra of the selected and reconstructed events in the TOP and BOT Ge detectors are compared over the full 20\,MeV$_\text{ee}$ range and in the region below 600 keV$_\text{ee}$. With no Pb shielding surrounding the cryostat, the rates were found to be significantly higher, ranging between 37-43 and 26-31\,counts/s for the BOT and TOP Ge detectors, respectively.

\begin{figure}[ht!]
    \centering
    \includegraphics[width=0.95\linewidth]{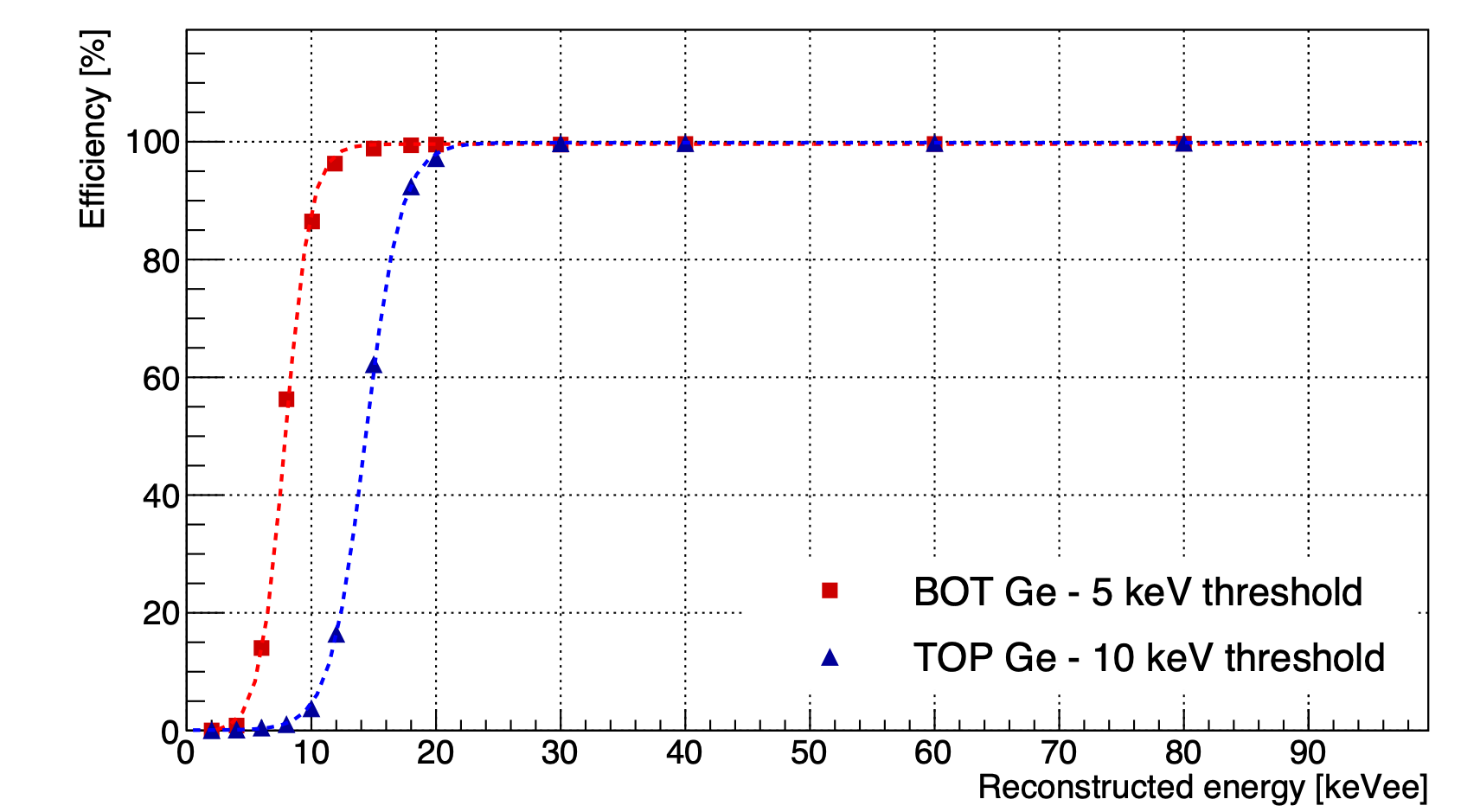}
    \captionsetup[figure]{width=\linewidth}
    \captionof{figure}{Pulse selection efficiencies for the HPGe detectors.
    } 
    \label{fig:RecEff_COV}
\end{figure}

The pulse selection efficiency was investigated by superimposing on the data, reference pulses with the same shape characteristics than those measured, and with amplitudes tuned for well defined energies. The analysis of the overlaid data thus allowed to determine the combined pulse selection and reconstruction efficiency as a function of energy. Figure \ref{fig:RecEff_COV} shows that an excellent efficiency, above 96$\%$, could be reached down to 12\,keV$_\text{ee}$ and 20\,keV$_\text{ee}$ for the Ge BOT and Ge TOP detectors, respectively.

As mentioned previously, good time resolution of the Ge veto detectors is necessary in order to limit accidental rejection of good candidates in the LWO bolometer. The arrival time of an event in the BOT and TOP Ge detectors was determined to better than 10\,$\mu$s, owing to the short rise time of pulses.~Given the counting rates, true coincidences between the two Ge detectors could clearly be observed with a negligible level of accidentals. This is illustrated also in Figure \ref{fig:Ge_ESpectra_COV}, where the TOP and BOT Ge energy spectra are also shown for events detected in coincidence (orange and green curves). With the exception of the $\mathrm{e^{+}e^{-}}$ 511 keV annihilation line, all other X-ray and gamma-ray lines corresponding to full absorption peaks disappear. Moreover, a large fraction of events detected in coincidence is located in the high energy portion of the spectra. These events come from atmospheric muons crossing both detectors. The rate of coincidences between the TOP and BOT Ge detectors for a typical background run using the Pb shielding is $\mathrm{\sim}$0.8 counts/s.  
\subsection{LWO bolometer results}\label{subsec:LWO_bolo}
\begin{figure*}[ht!]
    \centering
    \vspace{-0.5cm}   
    \includegraphics[width=0.85\textwidth]{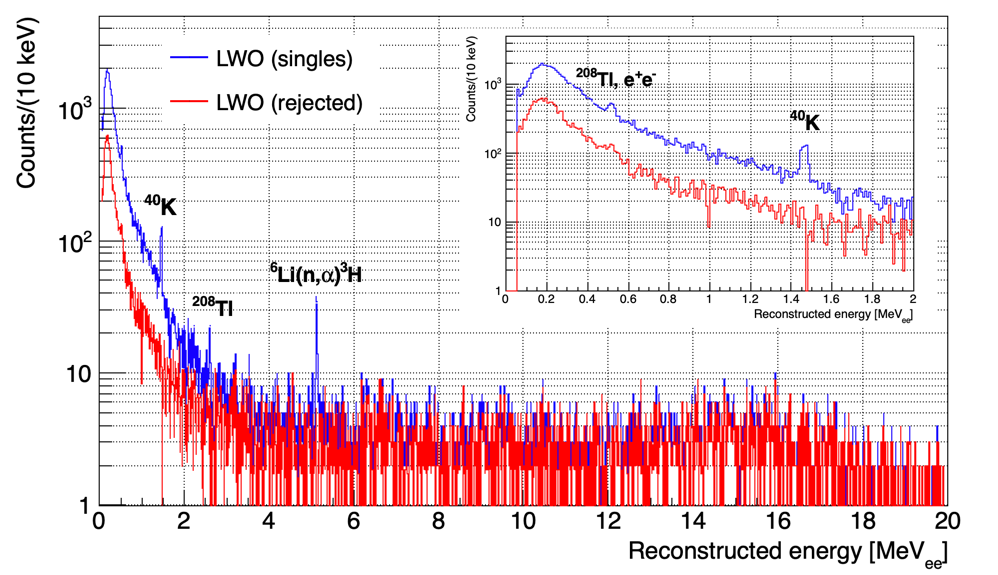}
    \captionsetup[figure]{width=\linewidth}
    \captionof{figure}{Energy spectrum of events detected in the LWO bolometer during a background run using the Pb shielding. The blue distribution corresponds to single events selected without any veto rejection. The red distribution corresponds to events selected in coincidence with at least one of the  HPGe ionization detector.}
    \label{fig:ESpectrum_NTD}
\end{figure*}

The calibration curve obtained by fitting a linear function to the identified gamma-ray peaks in the LWO detector is shown in Figure \ref{fig:Ge_Calib}(a). Taking into account a gain of 8000 from the NTD signal warm amplification stage, the sensitivity of the LWO bolometer was determined to be $\mathrm{\sim3}$\,nV/keV$_\text{ee}$. Non-linearities in the energy scale below 3\,MeV$_\text{ee}$ were found to be negligible. Figure \ref{fig:ESpectrum_NTD} shows the energy spectrum measured with the LWO bolometer during a run performed with the Pb shielding.~In addition to environmental gamma-ray lines, a peak at $\mathrm{5.12\pm 0.02}$\,MeV$_\text{ee}$ due to the $\mathrm{^6Li(n,t)\alpha}$ neutron capture reaction is visible. The peak intensity is here enhanced by the use of a $\mathrm{^{252}Cf}$ neutron source. It is shifted by $\mathrm{+(7.11 \pm 0.03)\,\%}$ with respect to the 4.78 MeV total energy released in the reaction because of the so-called thermal quenching effect. This effect is caused by the reduced scintillation yield of alpha and tritron particles with respect to electrons or gamma rays, which consequently lead to larger heat signals. This effect is expected to be of the order of a few percents in Lithium-containing scintillating bolometers \cite{BASKET}.

The measured energy resolution below 6\,MeV$_\text{ee}$ is shown on figure \ref{fig:Ge_Calib} (b). It is well-described by the following relation \cite{Armengaud_2017, Augier_2022}: 
\begin{equation}
    \mathrm{\frac{\sigma}{E} = \sqrt{\frac{\sigma_0^2}{E^2} + \frac{p_1}{E}},}
    \label{eq:resolution_LWO}
\end{equation}
where the parameter $\mathrm{\sigma_0}$ denotes the baseline intrinsic noise and $\mathrm{p_1}$ is related to the statistical fluctuations in the number of phonons. A fit of equation \ref{eq:resolution_LWO} to the LWO measured energy resolution data points gives $\mathrm{\sigma_0 = 9.3\pm 1.3}$\,keV$_\text{ee}$ and $\mathrm{p_1 = 0.032\pm 0.004}$\,keV$_\text{ee}$ ($\mathrm{\chi^2}$/NDF = 0.3), showing that below 6\,MeV$_\text{ee}$ the baseline noise contribution dominates. The baseline noise was also characterized by selecting empty baselines extracted from 250-ms long data streams. The average r.m.s. value of the baseline fluctuations was measured to be 5.7\,keV$_\text{ee}$ allowing thus a 25\,keV$_\text{ee}$ energy threshold for the event triggering.~The corresponding total rate of events in the LWO bolometer was about 2 counts/s during runs using the Pb shielding. Without Pb shielding, this rate increased to $\mathrm{\sim}$5 counts/s.~Similarly to the HPGe detectors, the efficiency for event selection in the LWO detector was determined using template pulses overlaid to baselines extracted from the data. A pulse selection efficiency better than 90\% is obtained for energies larger than 200\,keV$_\text{ee}$.
\subsection{Anti-coincidence analysis}\label{subsec:anti-coinc_analysis}
\begin{figure}[ht!]
    \centering
    \includegraphics[width= 0.85\linewidth]{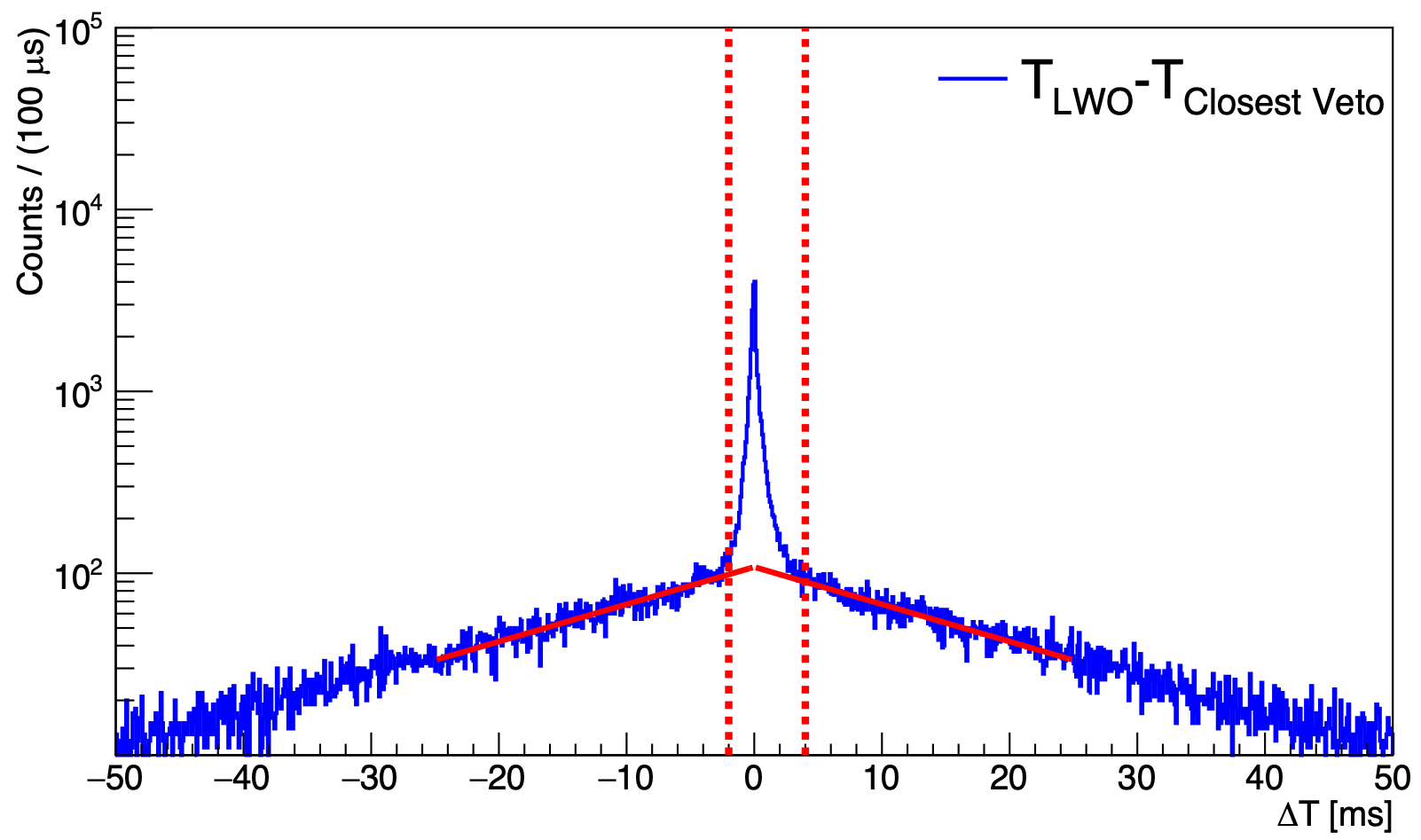}
    \captionsetup[figure]{width=\linewidth}
    \captionof{figure}{Distribution of time differences between LWO and HPGe selected events. The vertical dashed lines indicate the coincidence region while the red solid lines represent a fit to the side-band exponential tails.} 
    \label{fig:TClosest}
\end{figure}

The vetoing capability of the HPGe detectors was investigated by looking at time correlations with signals from the LWO bolometer. For the present setup, the rate of accidental coincidences is mainly dominated by the time resolution of the NTD sensor, which delivers much slower signals than the HPGe ionization detectors.~The time of an event in the LWO bolometer was estimated by the intersection between the baseline and a straight line extrapolated from the point where the slope of the pulse rise time is maximum (see figure \ref{fig:NTD_Pulses} b)). Using such a definition, figure \ref{fig:TClosest} shows the distribution of time differences between an event selected in the LWO bolometer and the closest event in time selected in any of the two HPGe detectors, after correcting for a 70-$\mathrm{\mu s}$ offset. As illustrated by the inset of figure \ref{fig:NTD_Pulses}, this offset likely sources from the LWO event time reconstruction method, which is systematically biased with respect to the true pulse onset time T$_0$. The corrected time difference distribution shows a clear peak centered at zero along with exponential tails at high $\mathrm{\Delta T}$ corresponding to accidental coincidences. The correlation peak features a full width at half maximum of $\mathrm{\sim 300\,\mu}$s. A time window (red dotted vertical lines on figure \ref{fig:TClosest}) is set from -2\,ms to +3\,ms to select true coincidence events. Bolometer events selected in coincidence with HPGe events with a time difference falling within this time-window were thus rejected. Given the relatively high counting rates observed in the HPGe detectors, the probability for accidental veto rejection turned out to be significant. In order to estimate the true veto rejection factor, a correction factor taking into account accidental rejection was applied. This correction factor was obtained by extrapolating the fitted exponential side tails into the correlation time window (see figure \ref{fig:TClosest}). For a typical run with Pb shielding, the fraction of LWO events in coincidence with at least one HPGe detector was found to be $\mathrm{39.6 \pm 0.2}$\% before correction and $\mathrm{32.2 \pm 0.2}$\% after correction. This result is in good agreement with the prediction of a simple toy MC that takes into account the measured rates of single and coincident events in the various detectors. Removing the Pb shielding, the fraction of rejected events by the HPGe veto detectors dropped at the 20\% level because of the increase in the environmental gamma background at low energy. 
The energy spectrum for LWO events rejected using this time-coincidence selection criterion is shown on Figure \ref{fig:ESpectrum_NTD} and is compared to the spectrum of single events. In order to determine the energy distribution for true coincidences, the LWO energy spectrum was corrected for by using events located outside the correlation time region, and normalized to the fraction of accidentally vetoed events.~As expected, no gamma-ray line is observed for the set of rejected events, except for the case of the 511 keV gamma-ray line originating from the annihilation of $\mathrm{e^{+}e^{-}}$ pairs. A similar conclusion can be drawn for events originating from neutron captures on $\mathrm{^{6}Li}$. Moreover, events in coincidence with the HPGe detectors account for most of the high-energy part of the single event spectrum, especially in the 6-20\,MeV$_{ee}$ region. These events correspond to atmospheric muons simultaneously crossing the detectors.

\section{Data to Monte Carlo comparison}\label{sec:Simulation}
\subsection{Simulation description}
A simulation of the experimental setup using the \Geant 10.7 library \cite{GEANT4} is performed to validate the experimental results and disentangle the different source contributions to the selected single event energy spectra. As shown on figure\,\ref{fig:COVProto}, a simplified geometry has been implemented, i.e. the detectors along with their mechanical housings, the main elements of the dry refrigerator and the surrounding Pb shielding.~Three main source contributions have been considered and separately simulated: (i) environmental gammas coming from natural radioactivity in the materials surrounding the setup, surface (ii) muons and (iii) neutrons both originating from the interactions of primary cosmic rays in the atmosphere. They are expected to be the most contributing in a setup operated at shallow depth.~Environmental gammas are isotropically shot towards the setup with an energy spectrum obtained from the simulation of the radioactive decay of $\mathrm{^{238}U}$, $\mathrm{^{232}Th}$ and $\mathrm{^{40}K}$ primordial radionuclides uniformly distributed in the concrete walls, ceiling and floor of a $\mathrm{25\,m^{2}}$ room. Atmospheric muon energy and angular distributions follow a modified Gaisser parametrization \cite{Tang_2006}.~Atmospheric neutrons are isotropically shot following an experimental spectrum as measured in \cite{Gordon_2005}. 

For this work, 10$^8$ primary muons and neutrons and 10$^9$ primary gammas have been generated, which corresponds to an equivalent acquisition time of 44 days and 31 days for neutrons and muons and 5 hours for environmental gammas, respectively. The so-called "Shielding" reference physics list released by the \Geant collaboration was used to model all particle interaction processes, except the electro-magnetic component for which the Livermore model was chosen instead of the standard one. For each simulated primary event, only the deposited energy in each of the three detectors is scored and saved for later analysis. As a first approximation, any time information, e.g. associated to the decay of activated radioisotopes or the deposition of energy into the active detectors, is disregarded. Therefore, only true detector coincidences generated over timescales possibly much longer than the acquisition run time are simulated.

The knowledge of the detector response is necessary to correctly compare data and simulated spectra. The deposited energy spectra output from the simulation are expressed in visible energy using the HPGe and LWO energy responses as detailed in section \ref{sec:Results}. The visible energy is randomly drawn from gaussian distributions $\mathcal{G}(E_\text{dep}, \sigma(E_\text{dep}))$ with $\sigma(E_\text{dep})$ following equations \ref{eq:resolution_HPGe} and \ref{eq:resolution_LWO} along with their respective best fit parameters. Thermal quenching effects, as mentioned in section \ref{subsec:LWO_bolo}, are also taken into account for the specific case of neutrons. At first order, each event energy in the LWO crystal is shifted by $\mathrm{+7\,\%}$ regardless of the details of the underlying energy deposition processes.

\subsection{Results}\label{sec:Results}
A multi-step procedure is followed to simultaneously fit the normalization of the simulated visible spectra to the experimental data from the HPGe and LWO detectors. It takes advantage of the fact that each source contribution strongly dominates in a particular energy regime. A simple $\mathrm{\chi^{2}}$, only taking into account statistical uncertainties from the experimental data and from the simulation, is computed and minimized to extract an estimate of each source contribution. The first step adjusts the atmospheric muon flux $\mathrm{\Phi_\mu}$ on single events with energies greater than 9 MeV. The second step uses the LWO $\mathrm{^6Li(n,t)\alpha}$ neutron capture peak region between 4.5 and 5.5 MeV to fit the flux $\mathrm{\Phi_n}$ of the atmospheric neutron component, including the previously fitted atmospheric muon component. The last step adds and fits the environmental gamma-ray flux $\mathrm{\Phi_{\gamma}}$ using the remaining low energy portion of the spectra between 0.5 and 3 MeV.

\begin{figure*}[ht!]
    \centering
    \includegraphics[width=0.95\textwidth]{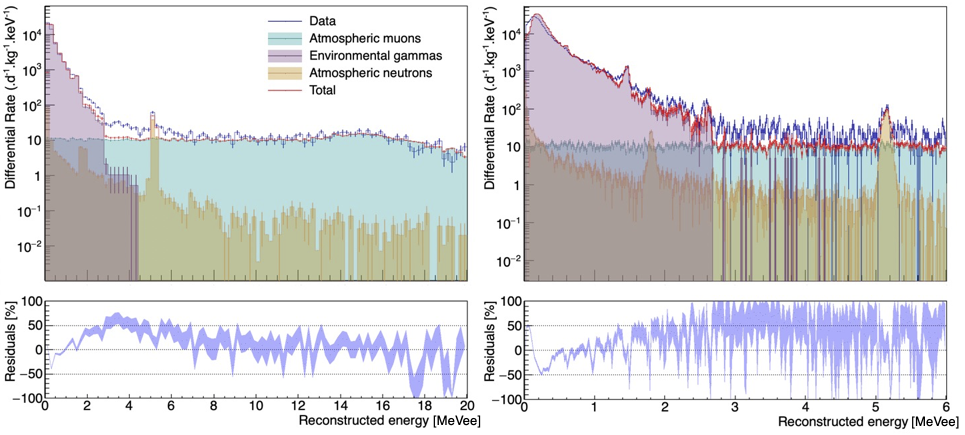}\captionsetup[figure]{width=0.95\linewidth}
    \captionof{figure}{Fit of the MC simulated data to the LWO single event spectrum as obtained from a 6-hour run. The left panel depicts the data to MC agreement over the full energy range up to 20 MeV, while the right panel focuses on the low energy portion below 6 MeV. The best fit flux normalization of respectively the environmental gamma (magenta-filled histogram), the atmospheric muon (cyan-filled histogram) and atmospheric neutron (orange-filled histogram) contributions are summarized in table \ref{tab:Fluxes}.}
    \label{fig:Fit_LWO}
\end{figure*}

\begin{figure}[ht!]
    \centering
    \includegraphics[width=0.95\linewidth]{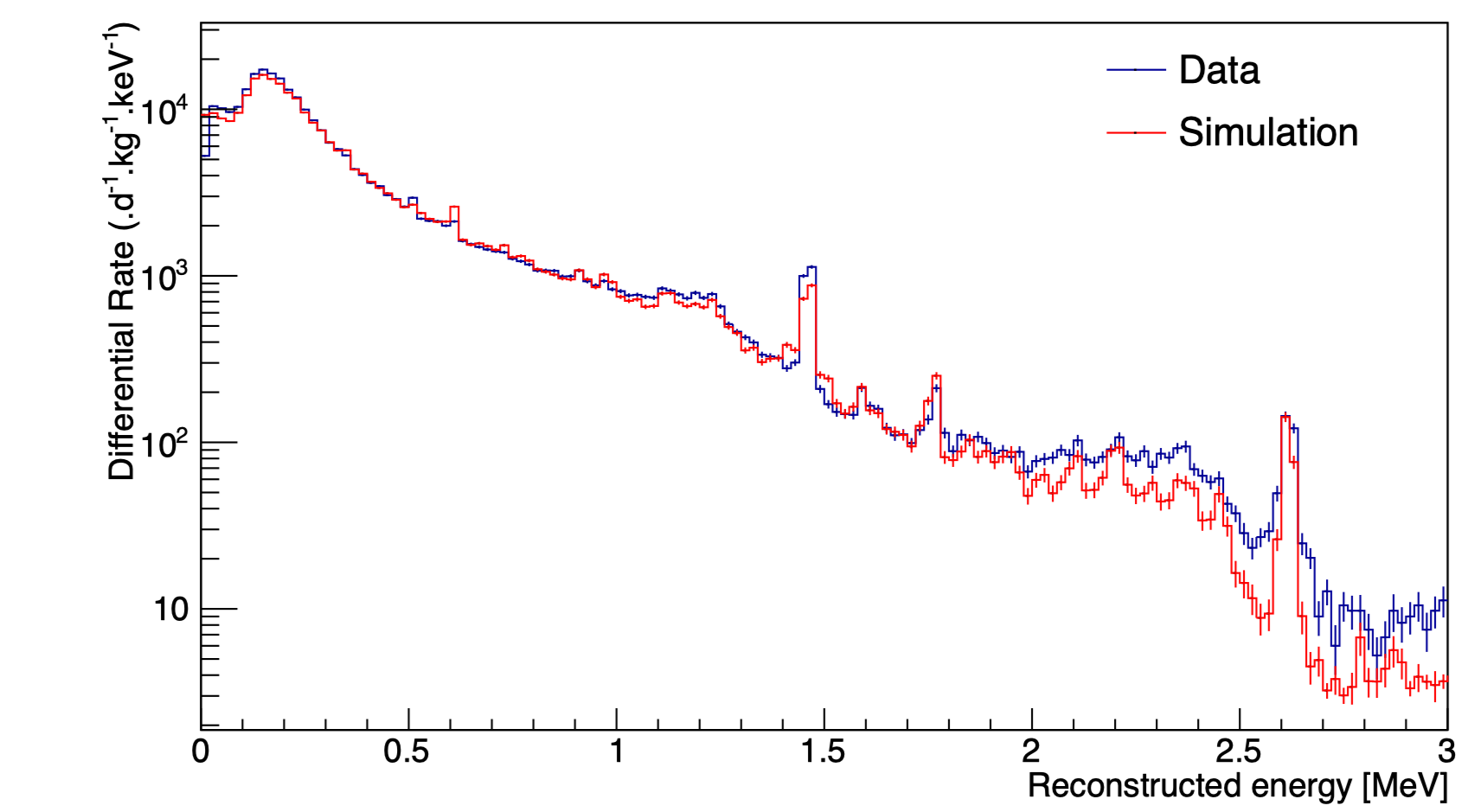}
    \captionsetup[figure]{width=\linewidth}
    \captionof{figure}{Single event energy spectrum (blue histogram) compared to the best fit MC simulation (red histogram) for the TOP HPGe detector in the environmental gamma-dominated region (0-3 MeV).} 
   \label{fig:Fit_Ge}
\end{figure}

Figure \ref{fig:Fit_LWO} shows the result of such a fitting procedure as obtained for the LWO detector and for a 6-h acquisition run taken without using any radioactive sources in the experimental setup and without lead shielding. The corresponding best fit flux values of the environmental gamma, atmospheric muon and atmospheric neutron contributions are reported in table \ref{tab:Fluxes}. They compare very well with the quoted literature values, which come from measurements carried out at surface or in very shallow laboratories. It should be noted that the environmental gamma ambiance has been independently measured at IJCLab using a liquid N2-cooled portable HPGe detector (Canberra GR3018). Three measurements over 5 days were performed in the room leading to a mean flux value of $\mathrm{3.2 \pm 0.3\, cm^{-2}\,s^{-1}}$ as quoted in table \ref{tab:Fluxes}.
The bottom panels of figure \ref{fig:Fit_LWO} depict the residuals of the best fit simulation with respect to the experimental data. The best-fit simulation compares fairly well to the data all along the full energy range, with especially a good agreement at the $\mathrm{\pm\,20}$\,\% level in the atmospheric muon-dominated portion of the spectrum above 8 MeV. Using a simple environmental gamma-ray generator based on the decay of primordial radionuclides in the concrete materials of an experimental room also reproduces fairly well the experimental data in the 0.05-3 MeV energy range. This agreement is even more remarkable when directly comparing the best fit simulation results with e.g. the TOP Ge data in that same energy range, as shown by figure \ref{fig:Fit_Ge}. While these do not qualitatively change the conclusions of the present data to MC comparison, some localized discrepancies are however visible. The gamma continuum culmination point around 200 keV is for example less well reproduced for the LWO than for the HPGe detector. This region corresponds to the transition where gamma-rays, which predominantly interact through Compton scattering at higher energies, start to experience photo-electric absorption. The position of this transition energy strongly depends on the details of the geometry and of the materials both in the close vicinity and in the active volume. The observed disagreement could then well reflect some lacks or oversimplifications in the simulated LWO environment geometry. Another possible cause, which goes well beyond the scope of the present work, is event selection and reconstruction issues at low energies due to possible noise instabilities in the electronic chain.

\begin{table*}[ht!]
    \centering
    \renewcommand{\arraystretch}{1.2}
    \begin{tabularx}{0.85\linewidth}{@{}@{\extracolsep{\fill}} lll@{}}
        \toprule
        \multirow{2}*{\textbf{Background contribution}} & \multicolumn{2}{c}{\textbf{Fluxes (/cm$\mathbf{^{2}}$/s)}} \\ \cmidrule{2-3}
        & This work & Reference values \\ \midrule
         Atmospheric muons & $(1.79\pm0.02) \times10^{-2}$ & $(1.90\pm 0.12)\times10^{-2}$ from \cite{Tang_2006}\\
         Environmental gammas & $3.126\pm 0.005$ & $3.2 \pm 0.3 ^{\star}$ \\
         Atmospheric neutrons & $(1.37\pm0.27) \times 10^{-2}$& $1.34 \times 10^{-2}$ from \cite{Gordon_2005}\\ \bottomrule
    \end{tabularx}
    \captionsetup[table]{width=0.95\linewidth}
    \captionof{table}{Best fit values for the integrated fluxes of each simulated background contribution. Reported uncertainties are statistical only. \\
    $^{\star}$ Measured at IJCLab with a HPGe portable spectrometer \textit{(See text for details)}.}
    \label{tab:Fluxes}
\end{table*}

The simulated counting rate also exhibits a $\mathrm{\sim}$30 \% deficit  with respect to the LWO data in the 2-8 MeV portion of the spectrum.~Similar deficits are also observed in the HPGe Monte Carlo to data comparison.~Complementing the simulated geometry with e.g.~adding the aluminum cryostat support structure has been checked to yield negligible changes. No significant correlation between the muon angular origin and a 2-8\,MeV energy deposition has been observed,  rejecting a deficiency in the atmospheric muon generator at large zenith angles.~A preliminary investigation showed that this discrepancy most likely comes from the atmospheric neutron spectrum used as an input to the simulation, which would need an increase of the flux in the very high energy $\mathrm{\gtrsim}$ 10 MeV portion to be cleared up. Other possible causes, which are numerous and could not be all investigated, are unaccounted shadowing effects of the experimental room concrete materials, improper quenching of the atmospheric neutron contribution or an unknown additional source not simulated in the present study.

\begin{figure*}[ht!]
    \centering  
    \begin{minipage}{0.5\textwidth}
        \centering
        \includegraphics[width=1.0\textwidth]{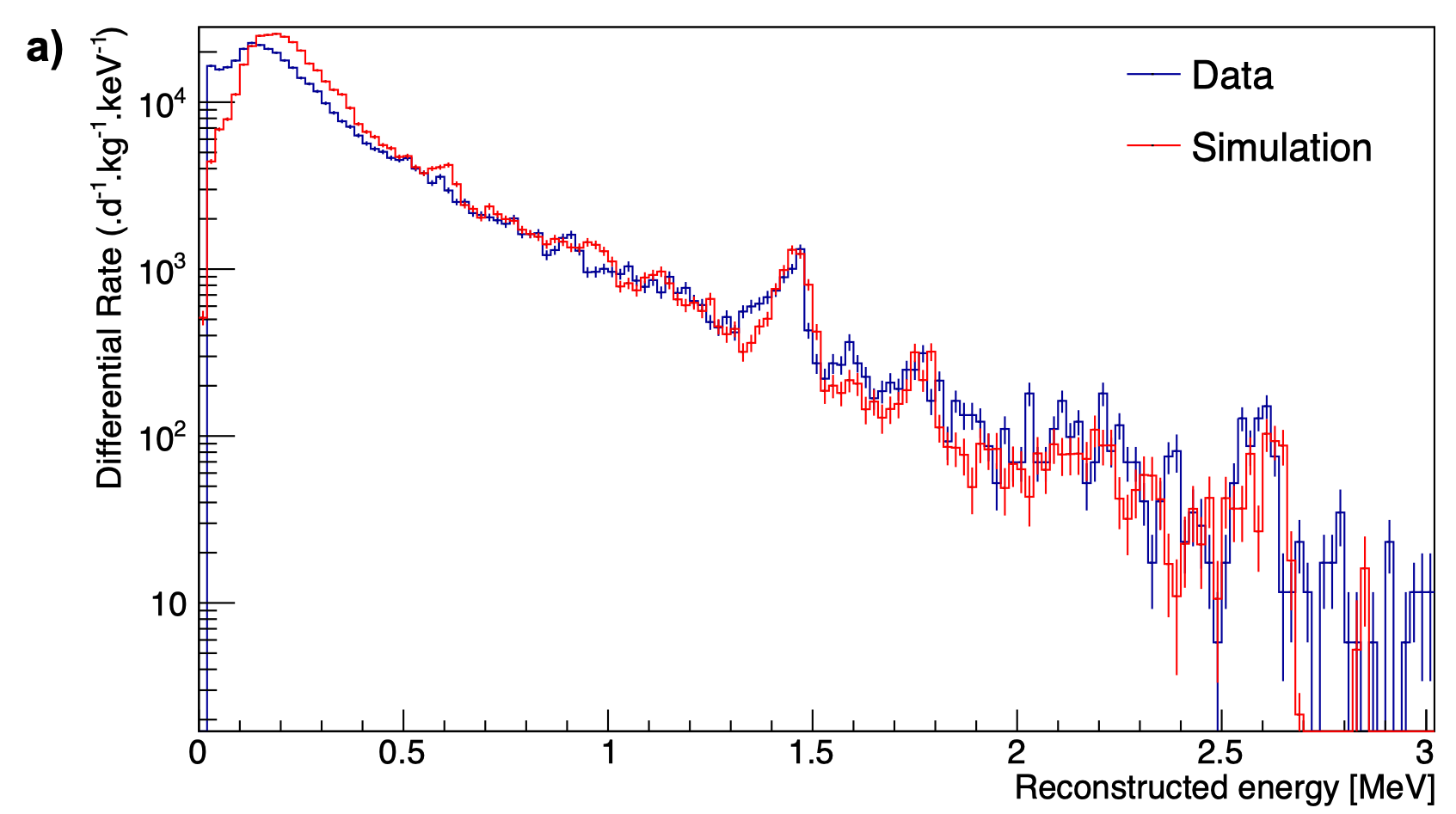}
    \end{minipage}\hfill
    \begin{minipage}{0.5\textwidth}
        \centering
        \includegraphics[width=1.0\textwidth]{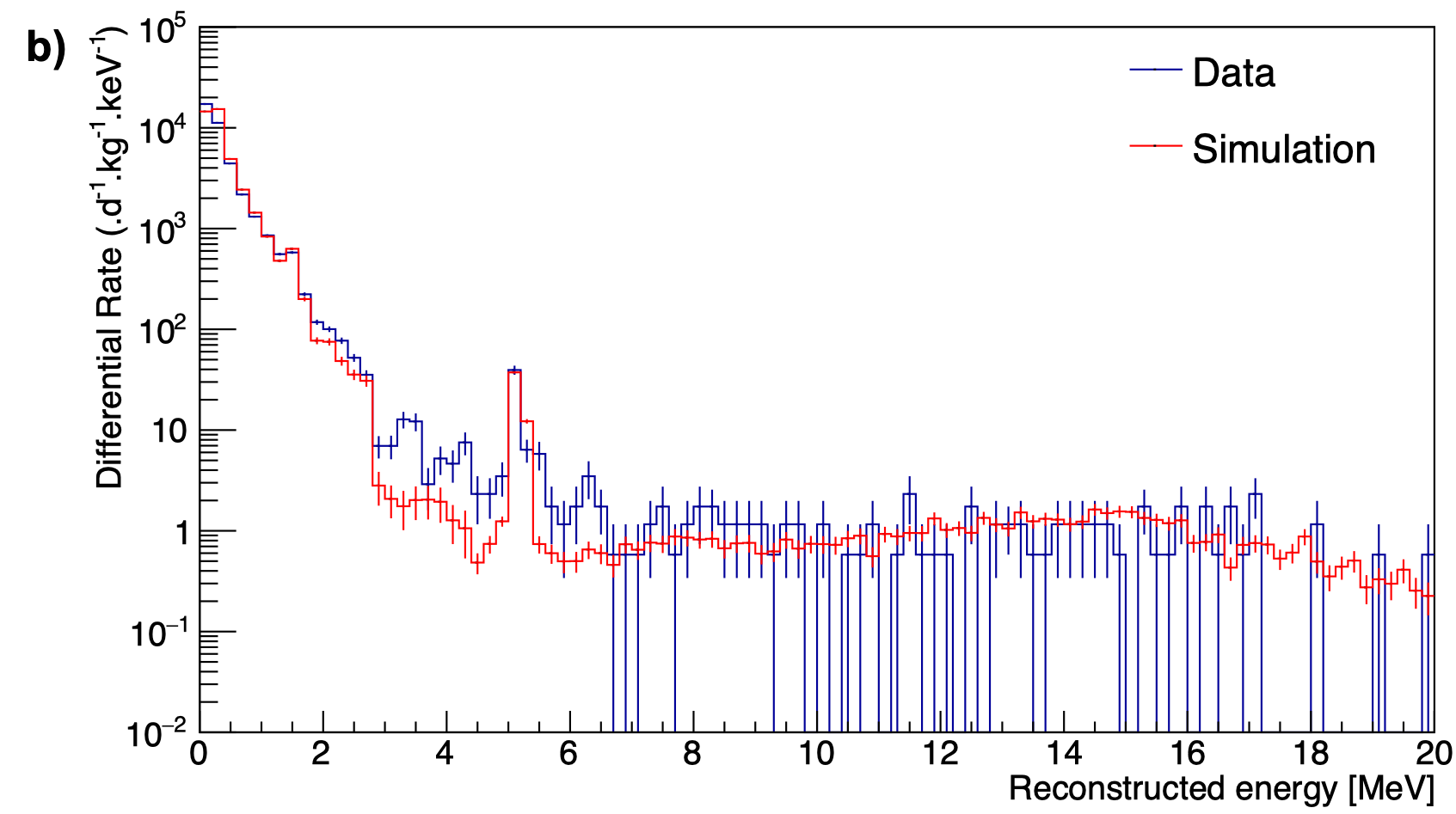}
    \end{minipage}
    \captionsetup[figure]{width=\linewidth}
    \captionof{figure}{Comparison of data and simulated spectrum of survived events in the LWO detector after applying an anti-coincidence selection criterion from the HPGe veto.}
    \label{fig:VetoEff}
\end{figure*}

The good understanding of each detector single event spectrum achieved by the present simulation work allows to test the anti-coincidence analysis and the LWO event rejection power. The fraction of vetoed events in the LWO detector is compared against the simulation prediction in table \ref{tab:VetoEff} for different energy ranges and with or without the use of the Pb shielding. Over the full energy range, a very good data-to-MC agreement is obtained. The ambient gamma rejection power in the 0.05-3 MeV energy regime is also well reproduced by the simulation.~The observed small deviation traces back to the incorrect reproduction of the LWO single event energy spectrum in the 200 keV energy region (see e.g. figure \ref{fig:Fit_LWO} (b)). This little deviation is also illustrated in the energy spectrum comparison of the LWO surviving events as shown by figure \ref{fig:VetoEff}. The rejection power to atmospheric muons, which can be appreciated looking at the fraction of vetoed events in the 10-20 MeV energy range, is equally well reproduced with a mean value of $\mathrm{\sim}$80 \%. Finally, the 3-10 MeV region shows the largest differences in the fraction of vetoed LWO events between data and simulation prediction.~As could be seen for the data-to-MC comparison of single event spectra, the fraction of atmospheric muon events and other sources of events is improperly simulated in this energy regime. The higher rejection power as predicted by the simulation can then either be related to an overestimate of the atmospheric muon contribution or an underestimate of other types of contribution, such as neutron or gamma-related events and which presents a less favorable veto rejection efficiency from the HPGe ionization detectors. A higher total rejection power is observed with the Pb shielding. This is mostly caused by a lower contribution from the environmental gamma events, which are less efficiently rejected as compared to the atmospheric muon events. Interestingly, the rejection power in the 3-10 MeV regime is unaffected with or without using the Pb shielding. An unaccounted source of external gamma-like events in the simulation seems then not to be responsible for the previously observed differences.

This overall good data-to-MC agreement finally allows to confidently extrapolate the performances of such a HPGe cryogenic veto system when extended to a $\mathrm{4 \pi}$ coverage, which can e.g. be achieved using an hermetic arrangement of many HPGe crystals. The background rejection performances of such a setup particularly apply to the \nucleus experiment, which currently develops a similar veto system. As anticipated, the gamma and atmospheric muon veto efficiency can respectively be improved up to 73.3\,\% and to 92.4\,\% when applying a 10 keV energy threshold as typically obtained for the HPGe detectors presented in this work. All sources of surface background considered, this then leads to a total rejection efficiency of 75.8\,\%, which drops to $\mathrm{\sim}$\,38\% when restrained to the sub-keV energy range, which is the energy range of interest for the CEvNS detection at a reactor. This decrease in the total rejection power comes from atmospheric neutrons, which are poorly rejected by the HPGe cryogenic veto and whose contribution dominates in this energy regime. \\
Lowering the energy threshold to 1\,keV slightly improves the total rejection efficiency to 74.8\,\% and 93.2\,\% for gamma-induced and muon-induced events respectively, leading to a total rejection efficiency of 77.9\,\% at all energies and to $\mathrm{\sim}$\,56\% at sub-keV energies. Achieving such low thresholds is more challenging, and would require for instance a careful design and integration of the electronic chain to mitigate and control unwanted sources of instrumental noise such as e.g. electromagnetic noise pickups and microphonics.

\begin{table*}[ht!]
    \centering
    \renewcommand{\arraystretch}{1.2}
    \begin{tabularx}{0.8\linewidth}{@{}@{\extracolsep{\fill}} lllll@{}}
        \toprule
        & \multicolumn{2}{c}{\textbf{No shielding}} & \multicolumn{2}{c}{\textbf{With shielding}} \\\cmidrule(lr){2-3} \cmidrule(lr){4-5}
        \textbf{Energy Range} & Data & Simulation & Data & Simulation \\\midrule
        $[0.05; 20]$ MeV & $22.0 \pm 0.5$\,\% & $22.4 \pm 0.4$\,\% & $30.8 \pm 0.5$\,\% & $30.1 \pm 0.7$\,\%\\
         $[0.05; 3]$ MeV & $20.4 \pm 0.5$\,\% & $22.0\pm0.4$\,\% & $26.3 \pm 0.5$\,\% & $28.8 \pm 0.8$\,\%\\
         $[3; 10]$ MeV & $93.1 \pm 3.4$\,\% & $84.2 \pm 0.9$\,\% & $93.2 \pm 2.0$\,\% & $84.4 \pm 2.0$\,\%\\
         $[10; 20]$ MeV & $78.4 \pm 3.3$\,\% & $81.7 \pm 1.3$\,\% & $80.2 \pm 2.1$\,\% & $81.3 \pm 2.2$\,\%\\
          $[50; 500]$ keV & $20.6 \pm 0.5$\,\% &  $23.1 \pm 0.5$\,\% & $25.6 \pm 0.6$\,\% & $29.7 \pm 0.8$\,\%
          \\\bottomrule
    \end{tabularx}
    \captionsetup[table]{width=0.95\linewidth}
    \captionof{table}{Rejection power measured in the data and in the simulation, with and without Pb shielding.}
    \label{tab:VetoEff}
\end{table*}

\section{Conclusions}\label{sec:conclusions}
This article reported about the operation of two HPGe ionisation detectors at cryogenic temperatures around a bolometric target detector in view of demonstrating the concept of a compact cryogenic veto system for reducing external particle backgrounds.~This background mitigation strategy is of particular interest for reactor \cevns{} experiments using cryogenic detectors.

Such a simple experimental apparatus already showed promising performances at a very low cost and without pushing for any optimization. An original and fast event reconstruction method, taking full advantage of the heat and ionisation pulse features, was developed to handle pile-up effects possibly arising in high background environments such as shallow depth or surface laboratories.~Using this method, 5-10 keV energy thresholds together with excellent event selection and reconstruction performances could be achieved for the HPGe ionisation detectors, first validating their suitability as veto detectors but also as powerful devices for characterizing the background environment in close vicinity to a target cryogenic detector. Although limited by the slow NTD response, an anti-coincidence analysis showed interesting gamma-like and muon-like event rejection powers at respectively the $\mathrm{\sim}$25\% and $\mathrm{\sim}$80\% levels.~Much higher background efficiencies could then be easily envisioned, by e.g. extending this veto system to a $\mathrm{4 \pi}$ coverage such as the one developed by the \nucleus{} collaboration.

Finally, a simple \Geant simulation of the experimental setup allowed to achieve a good description of the data, separating and measuring the main contributions at play in the energy spectrum of selected and reconstructed single events for each of the three detectors.~A multi step fit of the simulation to the data was used to infer the flux of environmental gamma-rays, atmospheric neutrons and atmospheric muons, and gave results very consistent with reported literature values at shallow depth laboratories. These measurements are of interest for modeling the expected particle backgrounds in shallow depth cryogenic detector experiments, for which the background environment is often not as well characterized as in deep underground laboratories. The good understanding of the particle source environment also allowed to reproduce with fairly good accuracy the rate and spectrum of the survived LWO events after applying anti-coincidence criteria with the HPGe detectors.~This allowed to confidently extrapolate the performances of such a cryogenic veto system when extended to a $\mathrm{4\,\pi}$ coverage, giving promising rejection powers at the $\mathrm{\gtrsim}$\,73\,\% and $\mathrm{\gtrsim}$\,92\,\% level for gamma-like and muon-like events, respectively.

\section*{Acknowledgements}
The authors acknowledge the financial support of the Cross-Disciplinary Program on Instrumentation and Detection of CEA, the French Alternative Energies and Atomic Energy Commission (BASKET Project). This work has also been partially funded by the P2IO LabEx (ANR-10-LABX-0038) in the framework of "Investissements d'Avenir" (ANR-11-IDEX-0003-01) managed by the French National Research Agency  (ANR).




\bibliographystyle{elsarticle-num} 
\bibliography{references}


\end{document}